\documentclass[doublecol]{epl2}
\usepackage{amsfonts,amssymb,amsmath,hyperref,graphicx,epsfig}
\usepackage{upgreek}
\usepackage{color}
\title{Estimating gravity acceleration from static atomic gravimeter by Kalman filtering}
\author{Bo-Nan Jiang\inst{1,2,3,4}\thanks{E-mail: \email{bnjiang@siom.ac.cn}} \and Yu-Zhu Wang\inst{5}}
\shortauthor{B.-N. Jiang}
\institute{
  \inst{1} School of Science, Shenyang University of Technology, Shenyang, Liaoning 110870, China\\
  \inst{2} Hefei National Research Center for Physical Sciences at the Microscale and Department of Modern Physics, University of Science and Technology of China, Hefei, Anhui 230026, China\\
  \inst{3} Shanghai Branch, CAS Center for Excellence in Quantum Information and Quantum Physics, University of Science and Technology of China, Shanghai 201315, China\\
  \inst{4} Shanghai Research Center for Quantum Sciences, Shanghai 201315, China\\
  \inst{5} Key Laboratory for Quantum Optics, Shanghai Institute of Optics and Fine Mechanics, Chinese Academy of Sciences, Shanghai 201800, China
}


\abstract{
We present the construction of the two-state model of the atomic gravimeter and the associated Kalman recursion to estimate gravity acceleration from atomic gravimeter. We find the Kalman estimator greatly improve the precision of estimates in short term by removing the white phase noise. The residual noise of the estimates follows 0.13 $ \upmu$Gal$/\sqrt{\rm{s}}$ for up to more than $100$ s and highlights a precision of 0.34 $\upmu$Gal at the measuring time of a single sample, even with no seismometer correction.}

\begin{document}
\maketitle
\section{Introduction}
Atom interferometry realizes a versatile tool that offers precise and accurate measurement for inertial sensing\cite{Kasevich,Geiger}, which greatly complements the state of the art classical instruments\cite{Ferier,Menoret,Wu,Bidel1,Bidel2,McGuirk,Caldani,Bertoldi,Durfee,Dutta}. Among these inertial sensors, atomic gravimeters\cite{Peters,Ferier,Menoret,Wu,Bidel1,Bidel2} are of great interest for a wide range of essential applications, from geophysics to fundamental physics\cite{Kai}.

When monitoring the waveform\cite{def:wf} of gravity acceleration, the atomic gravimeter reading fluctuates due to various measurement noises that comprises vibration noise, phase noise of the Raman lasers, detection noise, and ultimately the quantum projection noise\cite{Legouet}. Experimentally, after implementing a combination of vibration compensation\cite{Hensley,Merlot,Legouet}, phase locking\cite{Santarelli,Cacciapuoti}, and efficient detection scheme\cite{Rocco}, the residual measurement noise of atomic gravimetry is dominated by the white phase noise in the short term and demonstrates the corresponding $\tau^{-1/2}$ relations over measuring time $\tau$\cite{Ferier,Menoret,Wu,Hu,Gillot,Fu}. For static atomic gravimetry, the white phase noise is one of the most challenging obstacles we have to overcome when aiming to improve the short-term sensitivity\cite{Legouet}.

The central statistical problem of estimating gravity acceleration from static atomic gravimeter is obviously waveform estimation\cite{Brown}. Up to date, the solution most commonly used by the community is the average method, which applies long-term integration to achieve high precision at long measuring time, while leaving the short-term sensitivity unimproved\cite{Hu,Gillot,Ferier,Fu,Menoret}. However, since the average method depends on no statistical model of the atomic gravimeter and is non-real-time, some useful information of gravity variation would be distorted or filtered out, if the period of the phenomena is shorter than the measuring time. Therefore, a quasi-real-time estimation method rooted in the physics of the atomic gravimeter could be very promising to benefit the short-term performance and save the static gravimetry from the trade-off between precision and loss of information, since such a method not only significantly filters out the measurement noise, but also keeps the useful information of gravity variation in the short term.

For the waveform estimation problem of the system driven by white Gaussian processes and observed with white Gaussian noise, the Kalman recursion is an optimal estimator with the minimum mean square error, and provides quasi-real-time and causal estimation\cite{Kalman1,Kalman2}. Its application\cite{Vyazmin,Zheng,Zhang} to atom interferometry promises to benefit applications in inertial sensing\cite{Cheiney,Huang,Tennstedt}. Despite the promising expectations, a completely satisfactory implementation for static atomic gravimetry has not been found yet, since no works shed light on the statistical model of the static atomic gravimeter.

Thus, in this work, we present the construction of the two-state statistical model of the static atomic gravimeter and the associated Kalman recursion, and demonstrate a precise estimation (which is also intrinsically quasi-real-time) of the gravity acceleration. We find the Kalman estimator greatly improve the precision of estimates in short term by removing the white phase noise. The residual noise of the estimates follows 0.13 $ \upmu$Gal$/\sqrt{\rm{s}}$ for up to more than $100$ s and highlights a precision of 0.34 $\upmu$Gal at the measuring time of a single sample, even with no vibration compensation system. The reliability and robustness of the Kalman estimator are also confirmed.

\section{{\color{black}Two-state model of the atomic gravimeter}}
We start by a brief description of the apparatus. The atomic gravimeter is similar to the one described in Ref. \cite{Jiang}, except that we utilize no seismometer correction. The test mass is a free falling cloud of Rubidium 87 atoms. The atoms are initially trapped and cooled by a three dimensional magneto-optical trap to 3.7 $\mu$K. After state preparation and velocity selection, $N\sim 10^6$ atoms are selected and participate the interferometry that is composed of a sequence of three Raman pulses separated by two equal time intervals $T=82$ ms. After the interferometer sequence, we acquire transition probabilities from both the top ($0$) and each side ($\pm \frac{\pi}{2}$) of the central interference fringe (3 shots for each phase modulation), and determine the measured acceleration by the mid-fringe protocol. The direction of the momentum transfer $k_{\rm{eff}}$ of Raman transitions is reversed every 9 shots in order to reject the direction-independent systematic errors. Every sampling time $T_s=\rm{5.7 s}$, the interferometer cycles 18 shots and readouts one gravity acceleration $g(t)$. Comparing to the typical systems required to an atomic gravimeter\cite{Geiger}, this apparatus is distinguished by implementing no vibration compensation system. Instead, as shown below, the two-state statistical model is constructed to precisely estimate the gravity acceleration from noisy gravimeter readings.

The gravimeter reading fluctuates due to various measurement noises that comprises vibration noise, phase noise of the Raman lasers, detection noise, and ultimately the quantum projection noise\cite{Legouet}. The fundamental quantity used to characterize the noise on gravimetry is the residual acceleration, defined as the difference between the gravimeter reading and the true waveform of the gravity acceleration,
\begin{equation}
  y(t) = g(t)-g_0(t),
\end{equation}
where $y(t)$ is referred to as gravimeter noise, due to its stochastic nature.

Though describing the true state of the system in the waveform estimation problem, the state model does not resolve the state with unlimited precision because of the presence of some physical limits. Here, based on the physics of atom interferometry, the noise contribution to the state stems from the quantum projection noise. The quantum projection noise not only disturbs the state at each time $t$ by $w_1(t)$, but also generates phase error $w_2(t)$ between states at adjacent times. Therefore, the state model is built based on the noise characteristic originated in the quantum projection noise, and the residual acceleration can then be written as
\begin{equation}
  y(t) = w_1(t)+\xi_2(t),
\label{eq:y}
\end{equation}
where $w_1(t)\sim\mathcal{N}(0,Q_1)$ denotes a white Gaussian noise, and $\xi_2(t)$ an accumulated phase error that is responsible for the random walk nature of the residual acceleration,
\begin{equation}
  \xi_2(t) = \int^{t}_{0}w_2(t^{'})dt^{'},
\end{equation}
with the random process $w_2(t)\sim\mathcal{N}(0,Q_2)$ also being Gaussian.

We model the state of the gravimeter with a two-dimensional state vector
\begin{equation}
  X(t) =
  \begin{bmatrix}
  X_{1,1}(t) \\
  X_{2,1}(t)
  \end{bmatrix},
\label{eq:state vec}
\end{equation}
where we introduce auxiliary quantity $X_{1,1}(t)$ as an integration of the gravimeter reading
\begin{equation}
  X_{1,1}(t) = \int^{t}_{0}g(t^{'})dt^{'},
\end{equation}
and $X_{2,1}(t)=\xi_2(t)$. With $X(t)$, the two-state gravimeter model can be straightforwardly written as
\begin{equation}
  \frac{dX(t)}{dt} = F X(t) + B u(t) + W(t),
\label{eq:cont}
\end{equation}
where
\begin{equation}
  F = \begin{bmatrix}
  0 & 1 \\
  0 & 0
  \end{bmatrix} \nonumber, \quad
  B = \begin{bmatrix}
  1 \\
  0
  \end{bmatrix} \nonumber, \quad
  u(t) = g_0(t) \nonumber, \quad
  W(t) = \begin{bmatrix}
  w_1(t) \\
  w_2(t)
  \end{bmatrix} \nonumber.
\end{equation}
$F$ is the state transition matrix; $B$ is the control matrix; $u(t)$ is the control input; $W(t)$ represents the random process. Note that, though the gravity acceleration is time-varying due to gravimetric Earth tides, atmospheric and polar motion effects, and other phenomena, the periods of the waveform are much longer than the time interval between two readout\cite{Nibauer,Torge,Rothleitner}. Thus, we reasonably treat $g_0(t)$ as cyclostationary.

We now discretize Eq. \ref{eq:cont} based on integral approximation method and derive a discrete-time version of the two-state model along the time axis $t(n) = n T_s+t(0)$, which can be used to perform Kalman recursion
\begin{equation}
  X(n) = F_{\rm{d}}(T_s) X({n-1}) + B_{\rm{d}} u_{\rm{d}}(n) + W_{\rm{d}}(n),
\label{eq:discrete}
\end{equation}
where
\begin{equation}
  F_{\rm{d}}(T_s) = \begin{bmatrix}
  1 & T_s \\
  0 & 1
  \end{bmatrix} \nonumber, \quad
  B_{\rm{d}} = \begin{bmatrix}
  T_s \\
  0
  \end{bmatrix} \nonumber, \quad
  u_{\rm{d}}(n) = g_0(n) \nonumber,
\end{equation}
and
\begin{equation}
  W_{\rm{d}}(n)=\int^{t(n)}_{t(n-1)}F_{\rm{d}}[t(n)-t^{'}]W(t^{'})dt^{'},
\end{equation}
with the associated covariance matrix $Q$ given by
\begin{equation}
  Q = E[W_{\rm{d}}(n) W_{\rm{d}}(n)^T] = \begin{bmatrix}
  Q_1T_s+Q_2\frac{T_s^3}{3} & Q_2\frac{T_s^2}{2} \\
  Q_2\frac{T_s^2}{2} & Q_2T_s
  \end{bmatrix} \nonumber.
\end{equation}
The subscript "d" indicates the discrete-time version of the variable.

When observations are taken, the readout of the state is limited by the measurement noise. This gives an observation vector $Z(n)$, and a discrete-time observation equation of the form
\begin{equation}
  Z(n) = H X(n)  + V(n),
\label{eq:ob}
\end{equation}
where the measurement matrix $H$ gives the connection between the measurement and the state vector,
\begin{equation}
  H = \begin{bmatrix}
  1 & 0 \\
  \end{bmatrix} \nonumber,
\end{equation}
and $V(n)$ describes the noise contribution to the measurement of $X_{1,1}(n) = T_s\sum\limits^{n}_{m=0}g(m)$. Note that each gravimeter reading to be summed $g(m)$ makes an independent measurement and the noise contribution to $g(m)$ is Gaussian with zero mean and variance $R$. Thus, $V(n)$ also represents a Gaussian noise with zero mean and the variance defined by summing $R$ of $n+1$ independent $g(m)$, that is, $V(n)\sim\mathcal{N}(0,(n+1)RT_s^2)$.

In the framework of the two-state model, $X(n)$ is estimated by the Kalman recursion, and the estimate of gravity acceleration is then derived from the auxiliary quantity $X_{1,1}$ through the equation
\begin{equation}
  \widehat{g_0}(n) = \frac{\widehat{X}_{1,1}(n)-\widehat{X}_{1,1}({n-1})}{T_s}.
\end{equation}

Recently, the authors became aware of the works\cite{Galleani,Galleani2,Galleani3} on the two-state model of atomic clocks and its application to Kalman filtering. These works could be generalized to atomic gravimeters, after several significant modifications: First of all, for the state to be estimated, the noises considered and their contribution to the state are determined by the physical limit to resolving the state. Therefore, the model needs to be built based on the physics of atom interferometry, and have the ability to theoretically predict the precision limit of the estimation. Secondly, the model needs to be modified to properly describe the new subject, since in atomic gravimetry, the state to be estimated is time-varying and related to the absolute phase, but not phase fluctuations or frequency fluctuations.
\begin{figure}[tbp]
\includegraphics[width=0.5\textwidth]{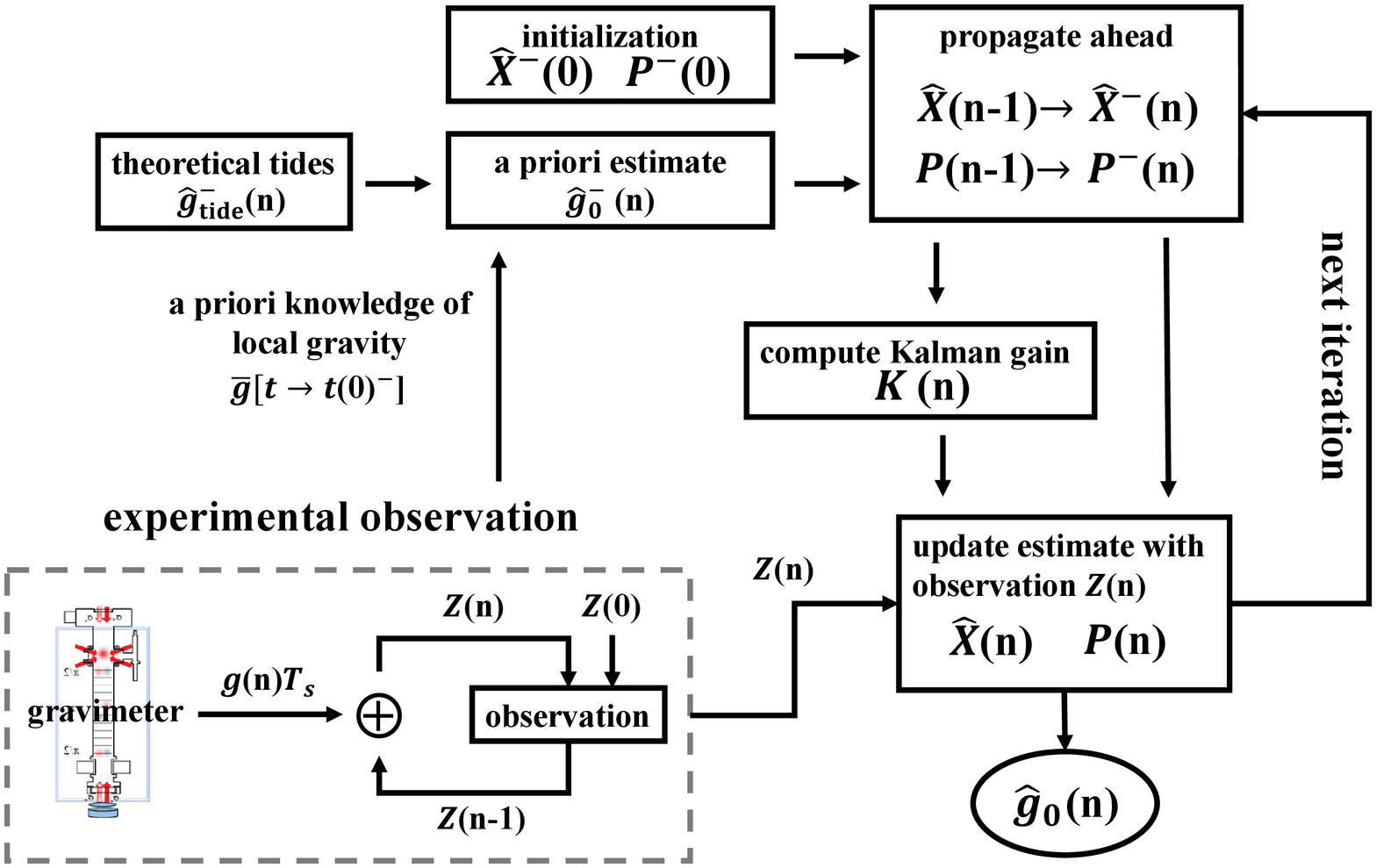}
\caption{(Color online) Kalman recursion for estimating gravity acceleration.}
\label{fig:Kalman}
\end{figure}

\section{The Kalman recursion}
\begin{figure*}[tbp]
\includegraphics[width=1.1\textwidth]{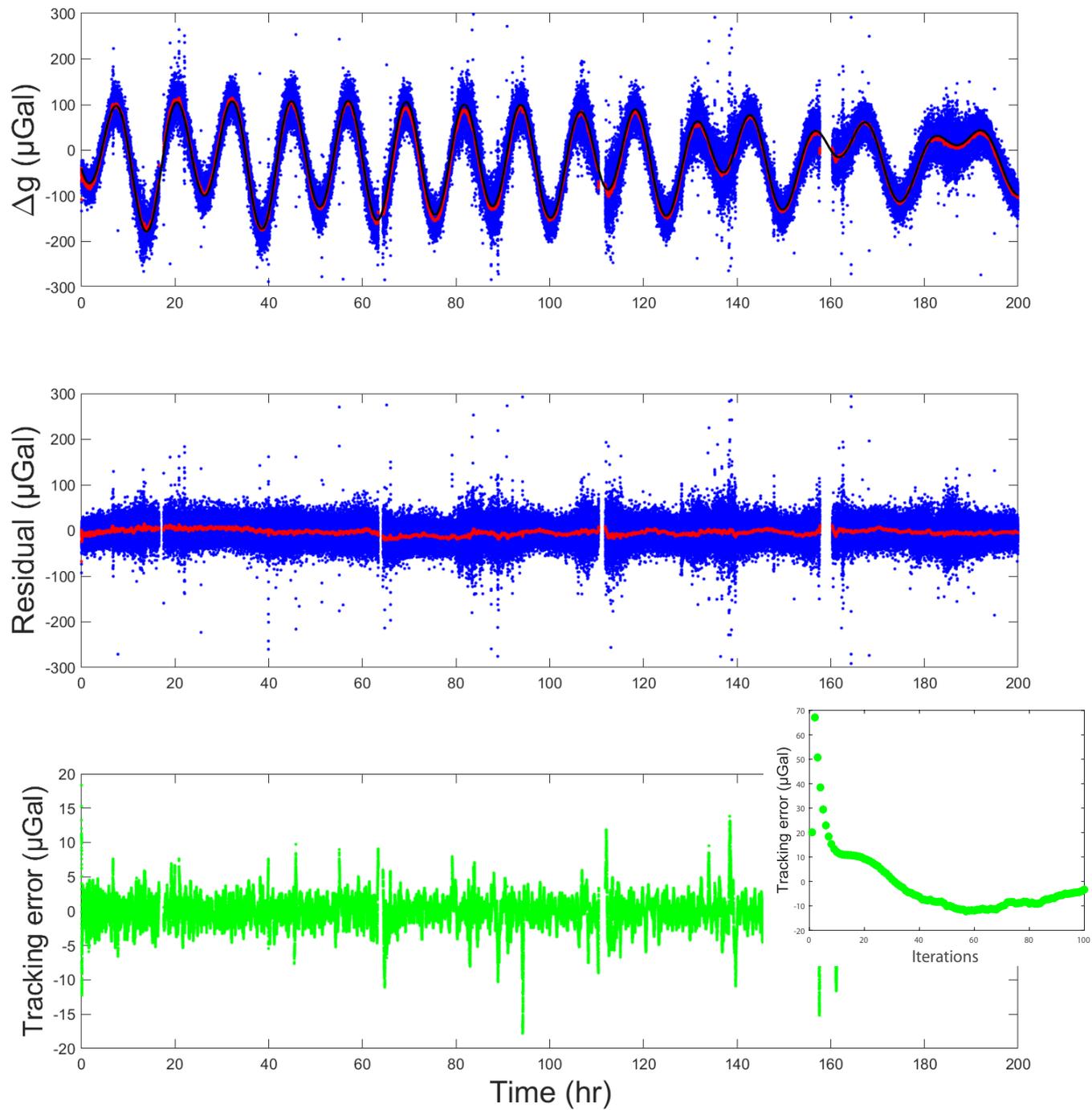}
\caption{(Color online) Long-term measurement of the local gravity for 200 hours and the corresponding residual acceleration by subtracting the gravimetric Earth tides from the gravimeter readings (blue) or the Kalman estimates (red). The tracking error (green) is obtained by comparing the estimates to the low-pass-filtered gravimeter output. And the black line is the theoretical prediction of the gravimetric Earth tides. The inset shows the tracking errors in the first 100 iterations.}
\label{fig:tide}
\end{figure*}

As shown in Fig. \ref{fig:Kalman}, we estimate the gravity acceleration from the gravimeter readings in a recursive mode, where the estimates $\widehat{X}(n)$, $\widehat{g_0}(n)$ and the associated error covariance matrix $P(n)=E\{[X(n)-\widehat{X}(n)][X(n)-\widehat{X}(n)]^T\}$ are constructed in two steps:

In the propagation step, $\widehat{X}^-(n)$, and $P^-(n)$ are predicted conditioned on the previous a posteriori estimates $\widehat{X}(n-1)$ and its covariance $P(n-1)$ as follows
\begin{eqnarray}
\widehat{X}^-(n) &=& F_{\rm{d}}(T_s) \widehat{X}({n-1}) + B_{\rm{d}} \widehat{g}_0^-(n), \nonumber \\
P^-(n) &=& F_{\rm{d}}(T_s)P({n-1})F_{\rm{d}}(T_s)^T + Q,
\label{eq:prop}
\end{eqnarray}
where the superscript "minus" indicates the a priori estimate deduced according to the two-state model of the atomic gravimeter. The a priori estimate $\widehat{g_0}^-(n)$ is theoretically generated through the equation
\begin{equation}
  \widehat{g_0}^-(n) =  \bar{g}[t\rightarrow t(0)^-] - \widehat{g}^-_{\rm{tide}}(0) + \widehat{g}^-_{\rm{tide}}(n),
\end{equation}
where $\bar{g}[t\rightarrow t(0)^-] = \lim\limits_{t\rightarrow t(0)^-}\frac{1}{\tau}\int^{t}_{t-\tau}g(t^{'})dt^{'}$ is an a prior measurement of the local gravity before $t(0)$, and $\widehat{g}^-_{\rm{tide}}(n)$ is the gravimetric Earth tides theoretically predicted with an inelastic non-hydrostatic Earth model\cite{Dehant}.

Then, in the update step, the a priori estimates $\widehat{X}^-(n)$ and $P^-(n)$ are improved after each observation $Z(n)$, and the a posteriori estimates $\widehat{X}(n)$, $\widehat{g_0}(n)$ and the covariance $P(n)$ are deduced according to
\begin{eqnarray}
\widehat{X}(n) &=& \widehat{X}^-(n) + K(n) [Z(n)-H\widehat{X}^-(n)], \nonumber \\
P(n) &=& [I-K(n)H]P^-(n), \nonumber \\
\widehat{g_0}(n) &=& \frac{\widehat{X}_{1,1}(n)-\widehat{X}_{1,1}({n-1})}{T_s},
\label{eq:update}
\end{eqnarray}
where the Kalman gain $K(n)$ that minimizes the mean square error is computed as
\begin{equation}
  K(n) = P^-(n)H^T [HP^-(n)H^T + (n+1)RT_s^2]^{-1},
\end{equation}
and to avoid growing memory per stack of data, the observation $Z(n)$ is constructed by summing the gravimeter readings $g(n)$ in a recursive procedure
\begin{equation}
  Z(n) = Z({n-1}) + g(n)T_s.
\end{equation}
Here, $Z(n)$ is experimentally obtained, but not through theoretical prediction $H\widehat{X}^-(n)$.

Note that, the parameters $Q_{1,2}$ (governs the covariance matrix $Q$) and $R$ of the two-state model affect the precision of the estimation. Here, $Q_{1,2}$ is determined based on the physics of atom interferometry, where the physical limit to determining the state of interferometry phase shift is quantum projection noise\cite{Tino}. For the white phase noise $y(n)\sim w_1(n)$, $w_1$ describes the quantum projection noise contribution to the state, and its variance can be written as
\begin{equation}
  Q_1=\left(\frac{1}{k_{\rm{eff}}T^2}\frac{1}{\sqrt{N}}\right)^2,
\end{equation}
and for the random walk phase noise $y(n)\sim \xi_2(n) = \int^{t(n)}_{0}w_2(t^{'})dt^{'}$, the integrand $w_2$ is also Gaussian and stems from phase errors between adjacent readings $g(n)-g({n-1})\sim w_2(n)T_s$, the variance of which is
\begin{equation}
  Q_2=\left(\frac{1}{k_{\rm{eff}}T^2}\frac{1}{\sqrt{N}}\right)^2\frac{1}{T^2_s},
\end{equation}
under the quantum projection limit. $R$ describes the measurement noise on gravimeter readings, and is experimentally determined from the variance of a sample of $g(n)$.

Based on the parameters determined above, the Allan deviation is theoretically calculated to be
\begin{equation}
  \sqrt{\frac{Q_1}{\tau}+\frac{Q_2\tau}{3}},
\end{equation}
where the crossover between the white phase noise and the random walk phase noise happens at $\tau_{\rm{x}}=\sqrt{\frac{3Q_1}{Q_2}}=9.8$ s. After $\tau_{\rm{x}}$, the estimator is dominated by the random walk phase noise $\sqrt{\frac{Q_2}{3}}=0.09 \upmu\rm{Gal}/\sqrt{\rm{s}}$ in the short term, which is the precision limit of the estimator. The precision limit could be treated as a systematic error that might bias the gravimetry at short term. However, for a measuring time as long as 100 s, this systematic error is well kept under 1 $\upmu\rm{Gal}$, which is a very acceptable level of accuracy\cite{Karcher}, even for a state-of-the-art atomic gravimeter.

The Kalman recursion is initialized according to our a prior knowledge about the atomic gravimeter and the local gravity, with
\begin{eqnarray}
\widehat{X}^-(0) &=& \begin{bmatrix}
  \bar{g}[t\rightarrow t(0)^-]T_s \\
  \sqrt{Q_2}T_s
  \end{bmatrix}, \nonumber \\
P^-(0) &=& Q, \nonumber \\
Z(0) &=& g(0)T_s.
\label{eq:init}
\end{eqnarray}

As demonstrated in the algorithm above, only one data, but not the whole data set of the gravimeter readings $g(n)$ is input at each iteration of the Kalman recursion. Though being applied to the post-measurement estimation in this work, this intrinsic characteristic of the algorithm makes it easy to be generalized to the quasi-real-time implementation\cite{Jiang2}.

\section{The data}
The experimental observation was carried out at a seismic station dedicated to seismic studies in Zhaotong city, Yunnan Province, China. The vibration noise of the observation site was estimated to be 89.8 $\upmu$Gal/shot, dominating the measurement noise. We located the atomic gravimeter at a gravity pillar without any seismometer correction, and performed a long-term measurement of the local gravity for 200 hours there. These data were chosen since they consisted of a fairly long record with sufficient quality. To test the robustness of the Kalman estimator, no preprocessing was necessary to remove outliers and gaps in the time series. The outliers are mainly due to the vibration noise, since we utilized no vibration compensation system. And the gaps are due to temporary failures of the power supply. These failures did not damage the gravimeter, and a remote restart was possible once electrical power was restored.

\section{Performance}
\begin{figure}[tbp]
\includegraphics[width=0.5\textwidth]{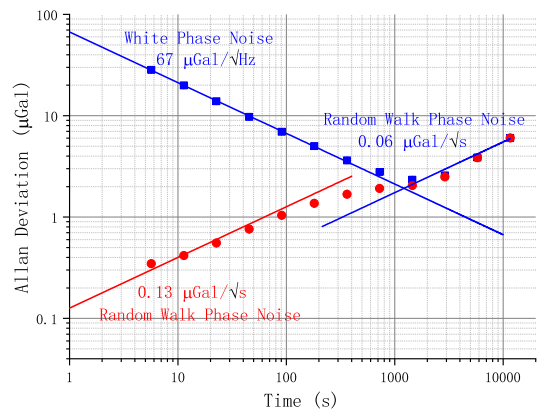}
\caption{(Color online) Allan deviation calculated from the residual acceleration in Fig. \ref{fig:tide}: blue square, the gravimeter readings; red dot, the estimates. The $\tau^{-1/2}$ and $\tau^{1/2}$ slopes represent the corresponding averaging expected for white phase noise and random walk phase noise. Since the Kalman estimator does not affect the long-term stability, the data points of the gravimeter readings and those of the estimates coincide after about 1000 senconds.}
\label{fig:allan}
\end{figure}

As shown in Fig. \ref{fig:tide}, both the gravimeter readings (blue) and the estimates (red) basically agree well with the theoretical prediction of the gravimetric Earth tides (black). The difference between the estimates and the theoretical prediction mainly reflects the effects of some non-solid-earth-tide phenomena, such as pressure or underground water changes. We also obtain the corresponding residual acceleration by subtracting the theoretical prediction of the gravimetric Earth tides from the gravity signal. The residual acceleration clearly shows that though the outliers and gaps exist in the gravimeter readings, their effects on the Kalman estimator are not significant. To evaluate the reliability of the Kalman estimator, we compare the estimates directly to the low-pass-filtered gravimeter readings. Note that the true waveform of the gravity acceleration are more complicated than the theoretical prediction of the gravimetric Earth tides, and is usually monitored by superconducting gravimeters during the comparison\cite{WuS}. Here, without superconducting gravimeters, the low-pass-filtered data serve as the true gravity acceleration, and the cut-off frequency of the low-pass filter is $\frac{1}{1450 \rm{s}}$. The tracking error of the Kalman estimator $\epsilon_{\rm{KF}}$ is obtained by subtracting the low-pass-filtered data $g_{\rm{LP}}$ from the estimates $\widehat{g_0}$,
\begin{equation}
  \epsilon_{\rm{KF}}=\widehat{g_0}-g_{\rm{LP}}.
\label{eq:trkerr}
\end{equation}
After about 15 iterations, the tracking error of the Kalman estimator stabilizes. The rms value of the tracking error is $\sim2.5$ $\upmu$Gal, which is in good agreement with the Allan deviation of the gravimeter readings with $\tau=1450$ s. The result shows that the tracking error is dominated by the noise of the low-pass-filtered data, but not the residual noise of the estimates. The Kalman estimator is thus more precise than the low-pass filter or the integration method used in previous works\cite{Hu,Gillot,Ferier,Fu,Menoret}, at least in short term. The reliability of the Kalman estimator has also been confirmed by the numerical simulations in Supplementary Materials\cite{Suppl}.

The Allan deviation of the residual acceleration is then calculated to characterize the short-term sensitivity and the long-term stability of the gravimeter readings and the estimates. As shown in Fig. \ref{fig:allan}, the sensitivity of the atomic gravimeter (blue) follows 67 $ \upmu$Gal$/\sqrt{\rm{Hz}}$ for up to 1000 s. The measurement noise is dominated by white phase noise in this regime, and demonstrates $\tau^{-1/2}$ characteristic of reduction via integrating in time. Beyond 1000 s, the stability is degraded due to the accumulated phase errors between adjacent readings that is responsible for the random walk phase noise 0.06 $ \upmu$Gal$/\sqrt{\rm{s}}$. The Kalman estimator significantly reshapes the residual noise by removing the white phase noise in short term, while leaving the long-term stability unchanged. The residual noise of the estimates (red) then shows $\tau^{1/2}$ integrating that corresponds to random walk phase noise, and follows 0.13 $ \upmu$Gal$/\sqrt{\rm{s}}$ for up to more than $100$ s. The Allan deviation observed is consistent with the theoretical prediction of the total noise in the short term $\sqrt{0.09^2+0.06^2}=0.11$ $\upmu$Gal$/\sqrt{\rm{s}}$, which is root-sum-squares of the precision limit of the estimator 0.09 $ \upmu$Gal$/\sqrt{\rm{s}}$ and the long-term stability observed 0.06 $ \upmu$Gal$/\sqrt{\rm{s}}$. Though not affecting the long-term stability, the Kalman estimator does greatly improve the precision of estimates in short term and highlights a precision of 0.34 $\upmu$Gal at the measuring time of a single sample ($T_s=5.7$ s), while in previous works\cite{Hu,Gillot,Ferier,Fu,Menoret}, the precision at this level costs both the sophisticated vibration compensation system and a measuring time of hundreds of samples to reach.

\section{Conclusion}
In conclusion, we shed light on the two-state statistical model that is rooted in the physics of the atomic gravimeter. In the model, the precision limit to estimation is theoretically predicted based on the quantum projection noise, which saves the estimation from risking the accuracy of gravimetry. The model we proposed makes the implementation of the Kalman recursion in the static atomic gravimetry possible, demonstrating a quasi-real-time and precise estimation. By comparing the estimates to the gravimeter readings derived with no vibration compensation system, the reliability and robustness of the Kalman estimator are confirmed. Through the Allan deviation of the residual acceleration, we find the Kalman estimator greatly improve the precision of estimates in short term by removing the white phase noise. The residual noise of the estimates follows 0.13 $ \upmu$Gal$/\sqrt{\rm{s}}$ for up to more than $100$ s and highlights a precision of 0.34 $\upmu$Gal at the measuring time of a single sample, which costs previous works both the sophisticated vibration compensation system and a measuring time of hundreds of samples to reach.

Though being applied to the post-measurement estimation in this work, the Kalman algorithm demonstrated is intrinsically quasi-real-time, and can be easily generalized to the quasi-real-time implementation\cite{Jiang2} and benefits the short-term gravimetry.

The estimation algorithm presented in this work provides us not only a precise and robust estimator for the gravity acceleration, but also an opportunity to simply the portable atomic gravimeter further, that is to say, an opportunity to perform precise gravity measurement in relatively quite locations without implementing any vibration compensation system. This demonstration would be of great interest for those applications involving static measurements of gravity, such as geophysics\cite{Kai}.

\acknowledgments
We thanks Prof. Shuai Chen and his team for the work in experimental observation. This work is funded by the Scientific Research Project of the Educational Department of Liaoning Province 2022 and the Youth Program of National Natural Science Foundation of China (Grant No. 11804019), and also partially supported by the National Key R\&D Program of China (Grant No. 2016YFA0301601), National Natural Science Foundation of China (Grant No. 11674301), Anhui Initiative in Quantum Information Technologies (Grant No. AHY120000), and Shanghai Municipal Science and Technology Major Project (Grant No. 2019SHZDZX01).


\clearpage
\onecolumn
\addtocounter{figure}{-3}
\renewcommand{\thefigure}{S\arabic{figure}}
\addtocounter{equation}{-19}
\renewcommand{\theequation}{S\arabic{equation}}
\makeatletter

\begin{center}
{\LARGE
Supplementary Materials for Estimating gravity acceleration from static atomic gravimeter by Kalman filtering}
\end{center}

\section{Numerical simulations}
In the numerical simulations, the time variation of the gravity acceleration is simulated by the theoretical prediction of the gravimetric Earth tides $\widehat{g}_{\rm{tide}}(n)$; the gravimeter readings $g(n)$ are generated by the two-state model in Eq. \ref{eq:discrete} of the main text; and then the tide is estimated from the simulated gravimeter readings by the Kalman recursion.

We verify the reliability of the Kalman estimator by investigating the tracking error to both tide and instrument with various measurement noises. The tracking error to tide $\epsilon_{\rm{tide}}$ is obtained by subtracting the theoretical prediction of the gravimetric Earth tides from the estimates $\widehat{g_0}$,
\begin{equation}
  \epsilon_{\rm{tide}}=\widehat{g_0}-\widehat{g}_{\rm{tide}},
\end{equation}
and we define the tracking error to instrument $\epsilon_{\rm{inst}}$ by Eq. \ref{eq:trkerr} of the main text as
\begin{equation}
  \epsilon_{\rm{inst}}=\widehat{g_0}-g_{\rm{LP}}.
\end{equation}
The cut-off frequency of the low-pass filter for $g_{\rm{LP}}$ is $\frac{1}{1450 \rm{s}}$.

\begin{figure*}[tbp]
\includegraphics[width=1\textwidth]{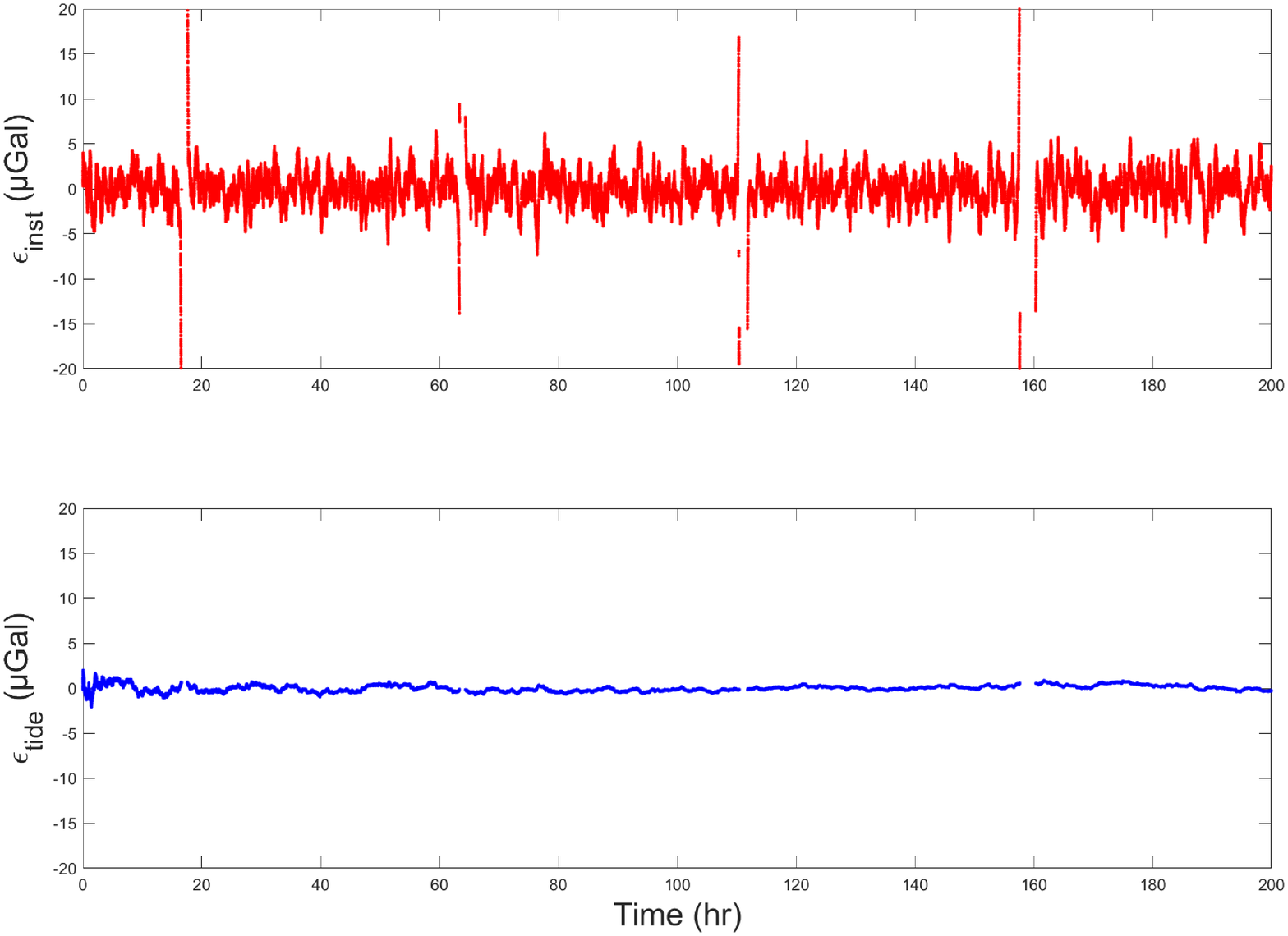}
\caption{(Color online) Numerical simulation with gravimeter readings fluctuating due to the white phase noise ($Q_1=67^2$ $\upmu\rm{Gal}^2/Hz$, $Q_2=0$ $\upmu\rm{Gal}^2/s$). Top: the tracking error to instrument. Bottom: the tracking error to tide.}
\label{fig:white}
\end{figure*}

In Fig. \ref{fig:white}, the simulated gravimeter readings fluctuate due to the white phase noise, whose variance is set to be $Q_1=67^2$ $\upmu\rm{Gal}^2/Hz$ according to the experimental observation in the main text. In this simulation, with $\epsilon_{\rm{tide}}$ and $\epsilon_{\rm{inst}}$ demonstrating zero mean, the Kalman estimates agree well with both the theoretical tide and the low-pass-filtered instrument output. The precision of the Kalman estimator is evaluated to be 0.23 $\upmu\rm{Gal}$ with the rms value of $\epsilon_{\rm{tide}}$. Though not every single Kalman estimate exactly falls on the theoretical tide, a deviation (or systematic error) below 1 $\upmu\rm{Gal}$ is very acceptable for a state-of-the-art atomic gravimeter. The rms value of $\epsilon_{\rm{inst}}$ is $2$ $\upmu$Gal, which is in good agreement with the theoretical predication of the two-state model $\sqrt{\frac{Q_1}{1450\rm{s}}}=1.8 \upmu\rm{Gal}$. The simulation is consistent with the experimental observation that the tracking error $\epsilon_{\rm{inst}}$ is dominated by the noise of the low-pass-filtered data, but not the residual noise of the estimates.

\begin{figure*}[tbp]
\includegraphics[width=1\textwidth]{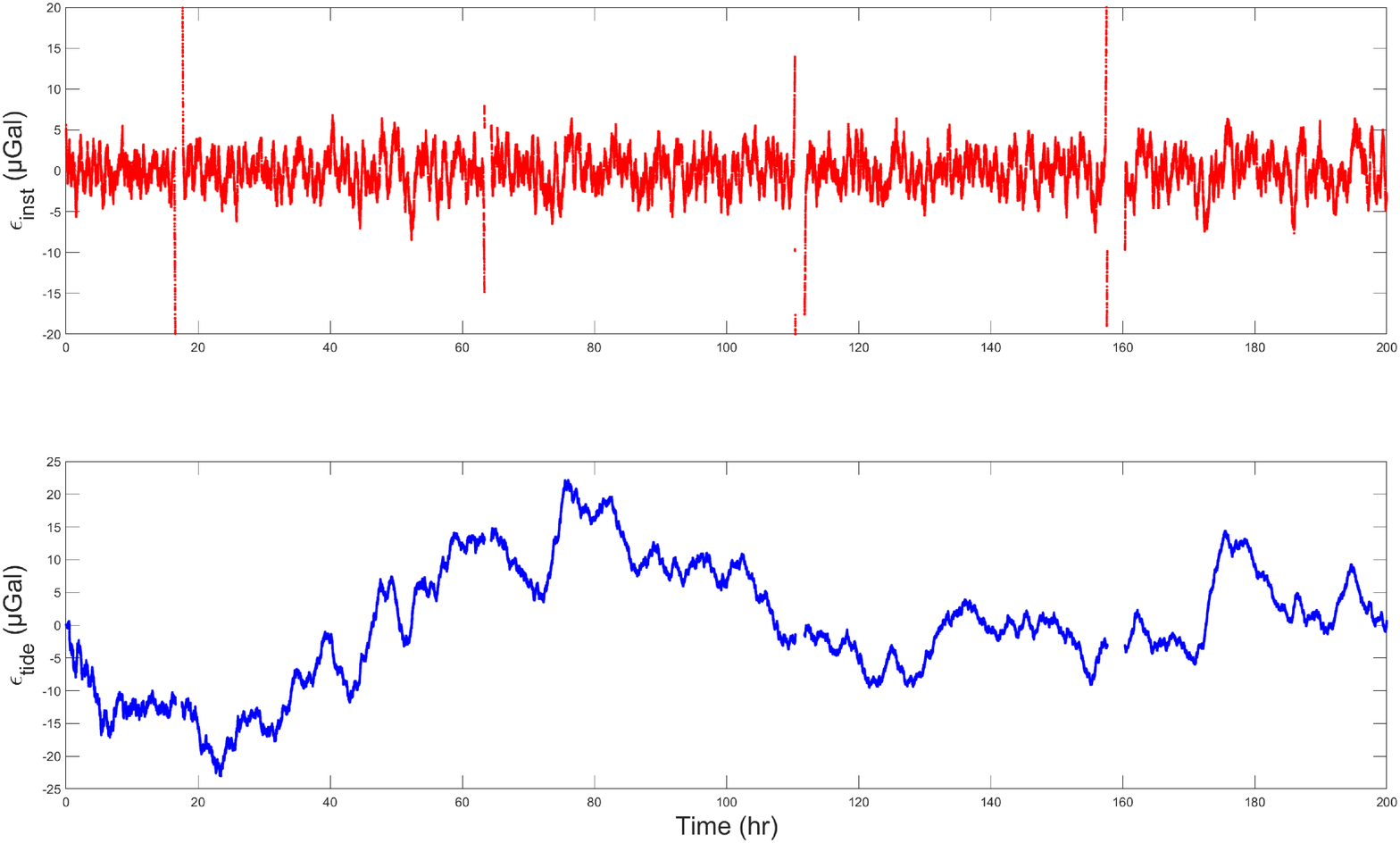}
\caption{(Color online) Numerical simulation with gravimeter readings fluctuating due to the white phase noise and the random walk phase noise ($Q_1=67^2$ $\upmu\rm{Gal}^2/Hz$, $Q_2=3\times0.06^2$ $\upmu\rm{Gal}^2/s$). Top: the tracking error to instrument. Bottom: the tracking error to tide.}
\label{fig:RW}
\end{figure*}

While in Fig. \ref{fig:RW}, both the white phase noise and the random walk phase noise play roles in the simulated gravimeter readings, and we set their variances to be $Q_1=67^2$ $\upmu\rm{Gal}^2/Hz$ and $Q_2=3\times0.06^2$ $\upmu\rm{Gal}^2/s$ according to the experimental observation in the main text. Comparing the behavior of $\epsilon_{\rm{tide}}$ and $\epsilon_{\rm{inst}}$, we find that though the tracking error to tide is zigzagged, the estimates still agree well with the low-pass-filtered instrument output, demonstrating zero mean and a rms value of 2.5 $\upmu\rm{Gal}$. It obviously shows that it is the noisy instrument, but not the Kalman estimator itself, that biases the estimates from the theoretical tide.

Comparing the results of the numerical simulations in Fig. \ref{fig:white} and Fig. \ref{fig:RW}, we find that the long-term stability of the atomic gravimeter is one of the key factors for improving the tracking error to gravity acceleration, since the Kalman estimator only significantly improve the short-term sensitivity, while leaving the long-term performance unchanged. However, as long as the Kalman estimator does not mathematically bias the gravimeter readings, it is a reliable method that will not obscure the phenomena we observe.



\begin{thebibliography}{}
\bibitem{def:wf} "waveform" stands for the time-varying state of a system, for example, the time-varying gravity acceleration which is to be estimated from the gravimeter readings.
\end{thebibliography}
\end{document}